\font\mysmall=cmr8
\def\btheta{\bar\theta}
\def\bkappa{\bar\kappa}

\def\itheta{{\mit\Theta}}
\def\idelta{{\mit\Delta}}
\def\igamma{{\mit\Gamma}}

\def\bR{{\rm I}\!{\rm R}}
\font\titlefont=cmbx10 scaled\magstep1
\magnification=\magstep1
\hfuzz 2pt

\null
\vskip 2cm
\centerline{\titlefont THE HIGGS PHENOMENON}
\smallskip
\centerline{\titlefont IN QUANTUM GRAVITY}
\vskip 2.5cm
\centerline{\bf R. Percacci}
\smallskip
\centerline{International School for Advanced Studies, Trieste, Italy}
\smallskip
\centerline{and}
\smallskip
\centerline{Istituto Nazionale di Fisica Nucleare, Sezione di Trieste}
\vskip 3cm
\centerline{\bf Abstract}
\smallskip\midinsert\narrower\narrower\noindent
The Higgs phenomenon occurs in theories of gravity in which the 
connection is an independent dynamical variable. The role of order 
parameters is played by the soldering form and a fiber metric. 
The breaking of the original gauge symmetry is linked to the 
appearance of geometrical structures on spacetime.
These facts suggest certain modifications and generalizations of the
theory. We propose a Higgs-like model which provides a dynamical 
explanation for
the nondegeneracy of the metric and a framework for the unification
of gravity with the other interactions.
\endinsert
\vskip 1.5cm
\centerline{Ref. S.I.S.S.A. 106/90/EP (July 1990)}
\vfil\eject

\beginsection 1. Introduction

One of the motivations for constructing a quantum theory of gravity 
is the hope that ultimately gravity will be unified with the other 
interactions. Since the electromagnetic, weak and strong nuclear forces 
are described by quantum theories, unification is not possible without 
quantization. 
The nongravitational interactions are all described by quantum
theories of a special type, namely quantum gauge theories.
The fact that General Relativity (GR) is also a gauge theory is encouraging
in this respect. However, the construction of unified theories including
gravity has used very different methods from those
used when gravity is left out: the unification of nongravitational
interactions is based on the Higgs mechanism, while most of the recent
work on the unification of gravity with the other interactions has been based
on higher dimensional theories.
The main aim of this paper is to motivate and present a framework for 
the unification of gravitational and nongravitational gauge interactions 
also based on the Higgs mechanism. A short presentation of these ideas 
has already appeared some times ago [1]. 

Here is a brief summary of the line of thought followed in this paper.
We will begin by showing that with certain
qualifications, a kind of Higgs mechanism is already operative in GR.
This is best seen by using a nonstandard formulation of GR which
makes use of many more fields than are strictly necessary.
Some of these fields are eliminated by constraint equations,
others by a larger than usual gauge group which includes, in addition to
coordinate transformations, also local $GL(4)$ transformations.
This formulation shows that, in the jargon of 
elementary particle physics,
GR is a ``spontaneously broken $GL(4)$ gauge theory'',
the role of order parameter being played by the metric and/or vierbein.
An ``unbroken'' phase of the theory would be characterized by a 
vanishing metric. 
This suggests that the most fundamental problem in 
quantum gravity is not to explain why the metric is curved instead of flat
but rather why the metric is nondegenerate instead of zero.

In elementary particle and condensed matter physics this kind of problems is
formulated and resolved within the context of the Higgs model.
In section 4, I will present a kind of mean field theory for gravity
based on a Higgs-like Lagrangian.
With certain reasonable assumptions about the positivity of the Hamiltonian,
it provides a selfconsistent dynamical explanation for the
nondegeneracy of the metric.

In section 5 I will present the promised 
model for the unification of gravitational and nongravitational
interactions. It is obtained from the model of section 4 
by enlarging the gauge group from $GL(4)$ to $GL(N)$. 
In a ``broken symmetry'' phase, characterized by a nondegenerate metric,
the $GL(N)$ gauge field can be split in a gravitational 
connection, an internal Yang-Mills field and additional mixed components, 
while in the ``symmetric'' phase, in which the metric vanishes,
no such splitting is possible. 
In the ``broken'' phase the Higgs phenomenon occurs and all components 
of the $GL(N)$ gauge field become massive, except for those 
corresponding to an $O(N-4)$ internal Yang-Mills field,
which can be thought of as the gauge field of a grand unified theory.
All this is in complete analogy to the unification schemes used for 
nongravitational interactions.

It will be clear from the whole discussion that the picture which emerges
will have definite implications on the problem of quantum gravity itself.
These will be briefly touched upon in section 6, together with
further comments, speculations and conclusions.

\beginsection 2. The $GL(4)$ formalism

In order to understand in what sense a Higgs phenomenon can be said to
occur in GR, it will be convenient to use a particular formalism which
exhibits invariance under local $GL(4)$ transformations [2,3]
(there are also different ways of seeing gravity as a spontaneously broken 
$GL(4)$ gauge theory [4]).  
Every theory of gravity can be presented in the $GL(4)$ formalism.
We review the relevant aspects here; a more complete
discussion in geometric language has been given in [3].
The $GL(4)$-invariant formulation of GR has also been studied in Hamiltonian 
form in a recent trilogy [5,6,7].

In the $GL(4)$ formalism, the independent dynamical variables describing
the gravitational field are an $\bR^4$-valued one-form $\theta^a{}_\mu$,
scalar fields $\kappa_{ab}$ in the symmetric tensor
representation of $GL(4)$ and, in first order formulations, 
a $GL(4)$ gauge field $A_\lambda{}^a{}_b$.
All indices run from one to four; however, greek indices refer to 
coordinate bases in the tangent spaces 
and latin indices refer to internal spaces, isomorphic to
$\bR^4$. These internal spaces form a vectorbundle $\xi_{\bR^4}$
over spacetime which is isomorphic to, but distinct from, the tangent
bundle $TM$. The one-form $\theta^a{}_\mu$ has to satisfy the constraint
$$\det\theta^a{}_\mu(x)\not=\, 0 \qquad\qquad\forall x\eqno(1)$$
and therefore describes an isomorphism from $TM$ to $\xi_{\bR^4}$.
We will call $\theta^a{}_\mu$ the soldering form.
The scalar fields $\kappa_{ab}$ are required to satisfy the constraint
$$\lambda_1(x)<0\ ,\ \lambda_2(x)>0\ ,\ \lambda_3(x)>0\ 
,\ \lambda_4(x)>0 \qquad\qquad\forall x \eqno(2)$$
where $\lambda_a$ are the eigenvalues of $\kappa$.
In particular, this implies $\det\kappa_{ab}(x)\not=\, 0$. 
Thus the fields $\kappa_{ab}$ describe a lorentzian fiber metric 
in $\xi_{\bR^4}$.
The metric $g_{\mu\nu}$ and the connection $A_\lambda{}^\mu{}_\nu$ 
in the tangent bundle $TM$ are obtained by pulling 
back $\kappa$ and $A$ by means of $\theta$:
$$g_{\mu\nu}=\ \ell^3\, \theta^a{}_\mu\, \theta^b{}_\nu\, \kappa_{ab}\ ,
\eqno(3)$$
$$A_\lambda{}^\mu{}_\nu=\theta^{-1}{}_a{}^\mu A_\lambda{}^a{}_b \theta^b{}_\nu
+\theta^{-1}_a{}^\mu \partial_\lambda \theta^a{}_\nu\ .\eqno(4)$$
Since $g_{\mu\nu}$ is assumed dimensionless and the dynamical fields
have canonical dimension of inverse length, we have to introduce in (3) a 
fundamental length $\ell$ (dimensional issues are further discussed in 
an Appendix, where we also show that $\ell$ has to be of the order of Planck's
length).
The gauge group is the group of linear automorphisms of $\xi_{\bR^4}$;
in local bases it can be described as consisting of
coordinate transformations $x^\prime=x^\prime(x)$ 
and local $GL(4)$ transformations $\Lambda(x)$. The fields transform as 
$$\eqalignno{\theta^a{}_\mu(x)&\mapsto{\theta^\prime}^a{}_\mu(x^\prime)= 
\Lambda^{-1 a}{}_b(x)\, \theta^b{}_\nu(x)
{\partial x^\nu\over\partial x^{\prime \mu}}\ ,&(5a)\cr
\kappa_{ab}(x)&\mapsto{\kappa^\prime}_{ab}(x^\prime)=
\Lambda^c{}_a(x)\, \Lambda^d{}_b(x)\, \kappa_{cd}(x)\ ,&(5b)\cr
A_\mu{}^a{}_b(x)&\mapsto A^\prime_\mu{}^a{}_b(x^\prime)=
{\partial x^\nu\over\partial x^{\prime \mu}}
\bigl(\Lambda^{-1 a}{}_c(x) A_\nu{}^c{}_d(x) \Lambda^d{}_b(x)+
\Lambda^{-1 a}{}_c(x)\partial_\nu\Lambda^c{}_b(x)\bigr)\ .&(5c)\cr}$$
From (5a) we see that it is possible to choose the $GL(4)$-gauge 
in such a way that $\theta^a{}_\mu=\ell^{-1}\delta^a_\mu$; in this gauge the
isomorphism between $TM$ and $\xi_{\bR^4}$ is fixed and thus one can 
identify the two bundles. It then follows from (3) that the dimensionless fiber
metric $\ell\kappa$ can be identified with the spacetime metric $g$. 
The local $GL(4)$ invariance is completely broken. 
The residual gauge group consists of transformations (5) with
$\Lambda^{-1 a}{}_b={\partial x^{\prime a}\over\partial x^b}$;
it is isomorphic to the group of coordinate transformations. In this gauge,
the theory reduces to the standard metric formulation.

On the other hand, from (5b) we see that it is possible to choose 
the $GL(4)$-gauge in such a way that $\kappa_{ab}=\ell^{-1}\eta_{ab}$,
where $\eta$ is the Minkowski metric.
In this gauge the fiber metric in $\xi_{\bR^4}$ 
is fixed. From (3) we see that in this gauge the
soldering form can be thought of as a vierbein. Furthermore,
the local $GL(4)$ invariance is broken to local Lorentz invariance.
In this gauge, the theory reduces to the vierbein formulation.

We can already see from this discussion that the fields $\theta$ and $\kappa$
behave in a certain sense like Higgs fields. Actually, the analogy is
even better if we compare them to a nonlinear sigma model (NSM) coupled to
gauge fields [3]. 
Recall that in the Higgs model the Higgs field $\Phi$ is a scalar field with 
values in a vectorspace $V$ carrying a representation of the gauge group $G$. 
The vectorspace $V$ can be decomposed into orbits of $G$ which in the
simplest cases have all the same stabilizer $H$ (with the exception of the
trivial orbit $\{0\}$). The Higgs field $\Phi$ can accordingly be
decomposed into a ``radius'' field $\rho$ and a $G/H$--valued ``angle''
field $\varphi$. A NSM is a Higgs field which is constrained to take
values in a fixed (nontrivial) orbit. In other words, the NSM field
is $\varphi$, with $\rho$ fixed. Gauge transformations act on $\varphi$
as follows: $\varphi(x)\rightarrow\varphi^\prime(x)=g^{-1}(x)\varphi(x)\ .$
Since $G$ acts transitively on $G/H$,
the field $\varphi$ has the property that it can be brought to a standard 
form (for example a constant) by means of a gauge transformation. 
The fields $\theta$ and $\kappa$ are similar to the field $\varphi$, 
since they also transform homogeneously and can be brought to a standard
form by means of a gauge transformation.
Actually, in the case of $\kappa$ this is more than just an analogy: 
since the space of lorentzian metrics in $\bR^4$ can be identified with the 
homogeneous space $GL(4)/O(1,3)$, the fiber metric $\kappa$ is literally a 
$GL(4)/O(1,3)$--valued nonlinear sigma model [3,8].

In the $GL(4)$ formalism, the torsion and the nonmetricity of the
connection $A$ are defined as follows.
The torsion of $A$ with respect to $\theta$ is 
$${\itheta}_\mu{}^a{}_\nu=
\partial_\mu \theta^a{}_\nu-\partial_\nu \theta^a{}_\mu+
A_\mu{}^a{}_b\, \theta^b{}_\nu-A_\nu{}^a{}_b\, 
\theta^b{}_\mu\ .\eqno(6)$$
The nonmetricity of $A$ with respect to $\kappa$ is
$$\idelta_{\lambda ab}=-\nabla_\lambda\kappa_{ab}= 
-\partial_\lambda \kappa_{ab}
+A_\lambda{}^c{}_a\, \kappa_{cb}
+A_\lambda{}^c{}_b\, \kappa_{ac}\ .\eqno(7)$$
$A$ is an irreducible
$GL(4)$ gauge field only if $\idelta\not=0$. In fact, if $\idelta=0$,
then in the vierbein gauge $A_{\lambda ab}$ is antisymmetric in the last
two indices and therefore reduces to an $O(1,3)$ gauge field.

Given $\theta$ and $\kappa$, there is a unique connection 
$\igamma(\theta,\kappa)$, 
called the Levi--Civita connection of $\theta$ and $\kappa$, 
which is torsionfree with respect to $\theta$ and
metric with respect to $\kappa$. 
Its components are $\igamma_\mu{}^a{}_b=\theta^d{}_\mu\,\kappa^{ac}
\, \igamma_{dcb}$, where
$${\mit\Gamma}_{abc}={1\over2}\bigl(
{\theta^{-1}}_c{}^\lambda\, \partial_\lambda\kappa_{ab}+
{\theta^{-1}}_a{}^\lambda\, \partial_\lambda\kappa_{bc}-
{\theta^{-1}}_b{}^\lambda\, \partial_\lambda\kappa_{ac}\bigr)
+{1\over2}\bigl(C_{abc}+C_{bac}-C_{cab}\bigr)\ ,\eqno(8)$$
and
$C_{abc}=\kappa_{ad}\, \theta^d{}_\lambda\bigl({\theta^{-1}}_b{}^\mu\,
\partial_\mu{\theta^{-1}}_c{}^\lambda-
{\theta^{-1}}_c{}^\mu\, \partial_\mu{\theta^{-1}}_b{}^\lambda\bigr)\ .$

\beginsection 3. General Relativity revisited

We now apply the general formalism of section 2 to the case of GR.
We want to recover Einstein's theory from a first order formalism,
so the action will contain the usual Palatini term
$$S_P(\theta,\kappa,A)={1\over16\pi G}\int d^4x\ \sqrt{|\det g|}\
\theta^{-1}{}_a{}^\mu\theta^b{}_\rho\, g^{\rho\nu} 
F_{\mu\nu}{}^a{}_b\ ,\eqno(9)$$
where $F_{\mu\nu}{}^a{}_b$ is the curvature of $A$. 
One can regard $S_P$ as the kinetic term for the gauge field. 
It is natural to add to the action covariant kinetic terms
for the fields $\theta$ and $\kappa$ as well. 
These will have the general form of squares of torsion and nonmetricity:
$$S_{\rm m}(\theta,\kappa,A)\!=\!\! \int \! d^4x \sqrt{|\det g|}
\left[A^\mu{}_a{}^{\nu\rho}{}_b{}^\sigma
\itheta_\mu{}^a{}_\nu\itheta_\rho{}^b{}_\sigma 
\! +\! B^{\mu ab\nu cd}\idelta_{\mu ab}\idelta_{\nu cd}
\!+\! C^\mu{}_a{}^{\nu\rho cd}\itheta_\mu{}^a{}_\nu\idelta_{\rho cd}\right]\! ,
\eqno(10)$$
where
$$\eqalignno{A^\mu{}_a{}^{\nu\rho}{}_b{}^\sigma=
&\ {1\over 2}\bigl( A_1\ell g^{\mu\rho}g^{\nu\sigma}\kappa_{ab}
+ A_2\ell^{-2}g^{\mu\rho} \theta^{-1}{}_b{}^\nu \theta^{-1}{}_a{}^\sigma
+ A_3\ell^{-2}g^{\mu\rho} \theta^{-1}{}_a{}^\nu \theta^{-1}{}_b{}^\sigma
\bigr)&(11a) \cr
B^{\mu ab\nu cd}=
&\ {1\over 2} \bigl(
  B_1 \ell^{-2}g^{\mu\nu}\kappa^{ac}\kappa^{bd}
+ B_2 \ell\, \theta^d{}_\rho g^{\rho\mu}
\theta^b{}_\sigma g^{\sigma\nu}\kappa^{ac}
+ B_3 \ell^{-2} g^{\mu\nu}\kappa^{ab}\kappa^{cd}\cr
&\qquad\qquad\qquad\quad 
+ B_4 \ell\, \theta^a{}_\rho g^{\rho\mu}
\theta^c{}_\sigma g^{\sigma\nu}\kappa^{bd}
+ B_5 \ell\, \theta^a{}_\rho g^{\rho\mu}
\theta^b{}_\sigma g^{\sigma\nu}\kappa^{cd}
\bigr) &(11b)\cr
C^\mu{}_a{}^{\nu\rho cd}=
&\ C_1 \ell\, g^{\mu\rho}\delta^c_a \theta^d{}_\sigma g^{\sigma\nu}
+ C_2 \ell^{-2}\theta^{-1}{}_a{}^\mu g^{\nu\rho}\kappa^{cd}
+C_3 \ell\, \theta^{-1}{}_a{}^\mu \theta^c{}_\sigma g^{\sigma\rho}
\theta^d{}_\lambda g^{\lambda\nu}\ .&(11c)\cr}$$
A factor $\ell$ has been inserted for each power of $\theta^a{}_\mu$
and $\kappa_{ab}$; in this way all coefficients $A_i$, $B_i$, $C_i$
are dimensionless. The reason behind the appearance of these factors is 
further discussed in the Appendix.

When the total action $S=S_P+S_{\rm m}$ is varied with respect to $A$ one gets 
(for almost every choice of the coefficients in (11)) the equation
$A_\mu{}^a{}_b=\igamma_\mu{}^a{}_b$, where $\igamma(\theta,\kappa)$,
is the Levi-Civita connection (8).
In order to prove this it is convenient to change variables from
$\theta$, $\kappa$, $A$ to $\theta$, $\kappa$, $\phi$, where
$\phi$ is a one-form with values in the Lie algebra of $GL(4)$
defined by
$$A_\lambda{}^a{}_b=\igamma_\lambda{}^a{}_b+\phi_\lambda{}^a{}_b\ .\eqno(12)$$
It transforms homogeneously under local $GL(4)$ transformations.
We have
$$\eqalignno{\itheta_\mu{}^a{}_\nu=&\ 
\phi_\mu{}^a{}_b\theta^b{}_\nu-\phi_\nu{}^a{}_b\theta^b{}_\mu\ ,&(13a)\cr
\idelta_{\mu ab}=&\ 
\phi_\mu{}^c{}_b\kappa_{ca}+\phi_\mu{}^c{}_a\kappa_{cb}\ ,&(13b)\cr
F_{\mu\nu}{}^a{}_b=&\, R_{\mu\nu}{}^a{}_b+\tilde\nabla_\mu\phi_\nu{}^a{}_b-
\tilde\nabla_\nu\phi_\mu{}^a{}_b+\phi_\mu{}^a{}_c\,\phi_\nu{}^c{}_b-
\phi_\nu{}^a{}_c\,\phi_\mu{}^c{}_b\ , &(13c)}$$
where $R_{\mu\nu}{}^a{}_b$ is the curvature of the Levi-Civita connection and
$\tilde\nabla$ denotes the covariant derivative with respect to the
Levi-Civita connection). Up to a surface term,
the action $S=S_P+S_{\rm m}$ can be rewritten in the form
$$S(\theta,\kappa,A)=S(\theta,\kappa,\igamma+\phi)=S_H(\theta,\kappa)
+S_Q(\theta,\kappa,\phi)\ , \eqno(14)$$
where $S_H(\theta,\kappa)=S_P(\theta,\kappa,\igamma(\theta,\kappa))$ 
is the Hilbert action,
$$S_Q(\theta,\kappa,\phi)={1\over2}\ell^{-2}Q(\phi,\phi)
={1\over2}\ell^{-2} \int d^4x\ \sqrt{|\det g|}\  
Q^\mu{}_a{}^{b\nu}{}_c{}^d\, \phi_\mu{}^a{}_b\, \phi_\nu{}^c{}_d \ ,
\eqno(15)$$
and
$$\eqalign{Q^\mu{}_a{}^{b\nu}{}_c{}^d\, = &\  
q_1g^{\mu\nu}\kappa_{ac}\kappa^{bd}+ 
q_2g^{\mu\nu}\delta^b_c\delta^d_a+ 
q_3\ell^3\theta^d{}_\rho g^{\rho\mu}
\theta^b{}_\sigma g^{\sigma\nu}\kappa_{ac} 
+q_4\ell^{-3}\theta^{-1}{}_c{}^\mu\theta^{-1}{}_a{}^\nu \kappa^{bd} \cr
+&q_5\theta^d{}_\rho g^{\rho\mu}\theta^{-1}{}_a{}^\nu\delta^b_c
+ q_6 g^{\mu\nu}\delta^b_a\delta^d_c
+ q_7 \ell^3\theta^b{}_\rho g^{\rho\mu}
\theta^d{}_\sigma g^{\sigma\nu}\kappa_{ac}
+ q_8 \ell^{-3}\theta^{-1}{}_a{}^\mu\theta^{-1}{}_c{}^\nu \kappa^{bd} \cr
+&q_9\delta^b_a\theta^{-1}{}_c{}^\mu\theta^d{}_\sigma g^{\sigma\nu}
+q_{10} \theta^b{}_\rho g^{\rho\mu}\theta^{-1}{}_c{}^\nu\delta^d_a
+q_{11}\theta^{-1}{}_a{}^\mu \theta^b{}_\sigma g^{\sigma\nu}\delta^d_c\ .\cr} 
\eqno(16)$$
The dimensionless coefficients $q_i$ are related to $A_i$, $B_i$ and $C_i$ by
an invertible linear relation which is given explicitly in equation (A.5)
of the Appendix.
For almost every choice of the coefficients, the quadratic form 
$Q$ is nondegenerate. 
We assume in the following that this is the case. 
Then, varying the action with respect to $\phi$ just gives the equation 
$\phi=\, 0$, which is equivalent to the statement that $A=\igamma$,
or $\itheta=\, 0$ and $\idelta=\, 0$.
On the other hand, variation of $S$ with respect to $\theta$ and 
$\kappa$ yields Einstein's equations in vacuum.
Thus we see that in spite of the large number of variables
which are initially present, this reformulation of the theory is classically
entirely equivalent to Einstein's. 

Note that the addition of the term $S_{\rm m}$ to the action, which is
very natural in the $GL(4)$ formalism, allows us to obtain both
$\itheta=0$ and $\idelta=0$ as equations of motion.
This is impossible in the traditional Palatini formalism
with action $S=S_P$ [3,6].

In a certain sense, the Higgs phenomenon is already visible in 
equation (14). Since the field equations imply that $A=\igamma$, 
the field $\phi$ describes the deviation of the dynamical gauge field $A$
from its dynamically determined value $\igamma$. 
Then, the kinetic term (10) for the fields $\theta$ and
$\kappa$ becomes a mass term for the deviation field $\phi$.
In particle physics there is no analog of the Levi-Civita connection;
instead, the Higgs phenomenon is seen by expanding 
the fields around the ground state, which in the
``broken symmetry'' phase is given by $A=0$, $\Phi={\rm const}\not= 0$.
There is an analog of this procedure also in GR. With the appropriate
boundary conditions the ground state of GR is flat Minkowski space.
In a suitable gauge it is represented by
$$\theta^a{}_\mu=\ell^{-1}\delta^a_\mu \quad,\quad
\kappa_{ab}=\ell^{-1}\eta_{ab}\quad,\quad 
A_\lambda{}^a{}_b=0\ .\eqno(17)$$
When the action is expanded around this ground state, it contains a mass term
$$\eqalign{{1\over2}\ell^{-2}\int d^4x\ A_\mu{}^a{}_b A_\nu{}^c{}_d
\bigl[ & \,
 q_1\eta^{\mu\nu}\eta_{ac}\eta^{bd}
+q_2\eta^{\mu\nu}\delta^b_c\delta^d_a
+q_3\eta^{\mu d}\eta^{\nu b}\eta_{ac} \cr 
+&q_4\delta^\mu_c\delta^\nu_a\eta^{bd}
+q_5\eta^{\mu d}\delta^\nu_a\delta^b_c
+q_6\eta^{\mu\nu}\delta^b_a\delta^d_c
+q_7\eta^{\mu b}\eta^{\nu d}\eta_{ac} \cr
+&q_8\delta^\mu_a\delta^\nu_c\eta^{bd}
+q_9\delta^b_a\delta^\mu_c\eta^{\nu d}
+q_{10}\eta^{\mu b}\delta^\nu_c\delta^d_a
+q_{11}\delta^\mu_a\eta^{\nu b}\delta^d_c \bigr]\ .\cr}\eqno(18)$$
As observed before, for almost all choices of the parameters in (11)
the mass matrix will be nondegenerate.
If we had not added the terms quadratic in torsion and nonmetricity,
the mass matrix would be degenerate.
Note that the mass eigenvalues will typically be of the order of
Planck's mass.

There are two ways in which this gravitational Higgs phenomenon
differs from the Higgs phenomenon we are accustomed to in particle physics: 
first, there are no propagating particles which acquire physical mass and 
second, we have not encountered the analog of the ``symmetric'' phase, 
in which $\langle\Phi\rangle=0$. 
In the next section we will discuss these differences in more detail
and try to eliminate them. This will lead us to the construction of 
a Higgs-like model for gravity.

\beginsection 4. A gravitational Higgs model

One of the central features of the Higgs phenomenon in particle physics
is that the spin 1 fields acquire physical mass. 
On the other hand the theory outlined in the previous section
is completely equivalent to Einstein's and therefore the only propagating 
particle it describes is a massless spin 2 mode.
Even though the kinetic terms of
$\theta$ and $\kappa$ give rise to a mass term for the fluctuation
of the gauge fields, these do not propagate and it would be perhaps
inappropriate to conclude that a true Higgs phenomenon is occurring in GR.
However, it is easy to construct different theories of gravity
with more propagating particles, by adding to the action terms 
quadratic in the curvature.
In the presence of torsion and nonmetricity the most general term of 
this type would be of the form 
$$S_2(\theta,\kappa,A)=\int d^4x\ \sqrt{|\det g|}\,
G^{\mu\nu}{}_a{}^{b\rho\sigma}{}_c{}^d\,
F_{\mu\nu}{}^a{}_b \, F_{\rho\sigma}{}^c{}_d \ ,\eqno(19)$$
where
$$\eqalignno{G^{\mu\nu}&{}_a{}^{b\rho\sigma}{}_c{}^d=
G_1 g^{\mu\rho}g^{\nu\sigma}\kappa_{ac}\kappa^{bd}
+G_2 g^{\mu\rho}\theta^{-1}{}_c{}^\nu\delta^d_a
\theta^b{}_\lambda g^{\lambda\sigma}
+G_3 g^{\mu\rho}\theta^d{}_\lambda g^{\lambda\nu}
\theta^{-1}{}_a{}^\sigma\delta^b_c \cr
+&G_4 g^{\mu\rho}g^{\nu\sigma}\delta^d_a\delta^b_c
+G_5\ell^{-3}g^{\mu\rho}\theta^{-1}{}_c{}^\nu
\theta^{-1}{}_a{}^\sigma\kappa^{bd}
+G_6 \ell^3g^{\mu\rho}\theta^d{}_\lambda g^{\lambda\nu}
\theta^b{}_\tau g^{\tau\sigma}\kappa_{ac} \cr
+&G_7 \theta^{-1}{}_c{}^\mu\theta^d{}_\lambda g^{\lambda\nu}
\theta^{-1}{}_a{}^\rho \theta^b{}_\tau g^{\tau\sigma}
+G_8 \ell^{-3}\theta^{-1}{}_a{}^\mu g^{\nu\sigma}
\theta^{-1}{}_c{}^\rho\kappa^{bd}
+G_9 \theta^{-1}{}_a{}^\mu \theta^d{}_\lambda g^{\lambda\nu} 
\theta^b{}_\tau g^{\tau\sigma}\theta^{-1}{}_c{}^\rho \cr
+&G_{10} \ell^3\theta^b{}_\lambda g^{\lambda\mu}g^{\nu\sigma}
\kappa_{ac}\theta^d{}_\tau g^{\tau\rho}
+G_{11} \theta^b{}_\lambda g^{\lambda\mu}\theta^{-1}{}_c{}^\nu
\theta^{-1}{}_a{}^\sigma\theta^d{}_\tau g^{\tau\rho}
+G_{12} \ell^{-1}\theta^{-1}{}_a{}^\mu g^{\nu\sigma}
\delta^b_c\theta^d{}_\tau g^{\tau\rho} \cr
+&G_{13} \theta^{-1}{}_a{}^\mu\theta^{-1}{}_c{}^\nu
\theta^b{}_\lambda g^{\lambda\sigma}\theta^d{}_\tau g^{\tau\rho}
+G_{14} g^{\mu\sigma}\theta^d{}_\lambda g^{\lambda\nu}
\delta^b_a\theta^{-1}{}_c{}^\rho
+G_{15} g^{\mu\sigma}\theta^{-1}{}_c{}^\nu\delta^b_a
\theta^d{}_\lambda g^{\lambda\rho} \cr
+&G_{16} g^{\mu\rho}g^{\nu\sigma}\delta^b_a\delta^d_c
+G_{17} \theta^{-1}{}_a{}^\mu\theta^b{}_\lambda g^{\lambda\nu}
\theta^{-1}{}_c{}^\rho\theta^d{}_\tau g^{\tau\sigma}\ .
& (20)\cr}$$
In general the action $S_P+S_{\rm m}+S_2$ will describe propagating 
particles of spins 2, 1 and 0 and the masses of these particles 
can be traced to the Higgs phenomenon. 
The perturbative properties of this action have been studied
in the case $\idelta=0$ [9]. No comparable study has been made in the
case $\idelta\not=0$. In particular, it will be important to establish whether
there exist choices of the coefficients for which the energy is bounded 
from below. In the following we are going to assume that such choices exist.
Further motivation for studying this type of action was given in [10].

The other difference with the usual Higgs phenomenon lies in 
the fact that we seem to be able to describe only the ``ordered''
or ``broken symmetry'' phase. This is due to the constraints
(1) and (2) which make the fields $\theta$ and $\kappa$ really
more akin to nonlinear sigma models than Higgs fields.
If we want to be able to describe other phases of the theory we have to
allow the fields $\theta$ and $\kappa$ to take values also in other orbits.
Thus we have to relax the constraints (1) and (2). Without them,
the configuration spaces of the fields $\theta$ and $\kappa$
are linear spaces, as is the configuration space of the Higgs field.

If we relax the constraints (1) and (2) we need to find a dynamical 
reason to explain why they hold to good approximation at low energy.
This is equivalent to explaining why the metric (3) is nondegenerate.
I think this is one of the central problems for quantum gravity.
It has been discussed recently in the context of Ashtekar's
reformulation of GR, which admits $g_{\mu\nu}=0$ as a solution [11,12,7].
However in GR there seems to be no local dynamical mechanism to
favor a nondegenerate metric over the zero metric. The choice seems to be
dictated by boundary conditions [13]. What we want instead is an
action that drives (the vacuum expectation value of) the metric
to be nondegenerate.

In the ordinary Higgs phenomenon, the expectation value of the
Higgs field is determined by the shape of a gauge-invariant potential.
Can a similar picture be carried over to the case of gravity?

When we try to write an action for gravity in the generalized setting
in which the constraints (1) and (2) do not hold, we run into a serious
problem, because the kind of action we would like to write requires a 
nondegenerate metric to construct the volume element and to contract indices.
Like most problems in quantum gravity, this can be traced to the dual role
played by the fields.
On one hand, they are used as geometrical standards of angles and lengths, 
and this requires them to be nondegenerate. On the other, they are dynamical
variables and one cannot see why they could not become degenerate 
due to quantum fluctuations.
\footnote{$^\ast$}{\mysmall 
While here I make the theory linear by relaxing the constraints (1) and (2),
the authors of [14] have taken the complementary approach
of trying to reconcile the quantum theory with the nonlinearity
implicit in the constraints.}

To avoid such problems we will adopt the rather drastic solution of 
separating these conflicting roles. The dynamical variables will still be
$\theta$, $\kappa$ and $A$, but the role of geometric standards will be
assigned to background fields $\bar\theta$, $\bar\kappa$ (assumed
nondegenerate) and their composites $\bar g$ and 
$\bar\igamma=\igamma(\bar\theta,\bar\kappa)$.
However, the dynamical and geometrical fields will not be completely unrelated:
in the end we will self-consistently identify the background fields
with the vacuum expectation values of the dynamical fields.
This is very much in the spirit of a mean field theory.
The action is
$$S(\theta,\kappa,A;\btheta,\bkappa)=
\bar S_2(\theta,\kappa,A;\btheta,\bkappa)+
\bar S_{\rm m}(\theta,\kappa,A;\btheta,\bkappa)+
U(\theta,\kappa;\btheta,\bkappa) \eqno(21)$$
where $\bar S_2$ and $\bar S_{\rm m}$ have the same forms 
of $S_2$ and $S_{\rm m}$ given in
(19) and (10) except for the replacement of $g$ by $\bar g$ in the
volume element and the replacement of $\theta$, $\kappa$ and $g$ by
the respective background fields in the tensors $A$, $B$, $C$ and $G$
defined in equations (11) and (20).
The last term in the action is a potential 
$$\eqalign{U(&\theta,\kappa;\btheta,\bkappa)=
\int d^4x \sqrt{|\det \bar g|}\Bigl[
{1\over 2} \ell^{-2}\mu^2\bar\theta^{-1}{}_a{}^\nu
\bar\theta^{-1}{}_b{}^\mu\theta^a{}_\mu \theta^b{}_\nu \cr
&+{1\over 4}\ell^{-4}
\left(\lambda_1 \bar\theta^{-1}{}_a{}^\nu \bar\theta^{-1}{}_b{}^\mu
\bar\theta^{-1}{}_c{}^\sigma \bar\theta^{-1}{}_d{}^\rho
+\lambda_2 \bar\theta^{-1}{}_a{}^\sigma \bar\theta^{-1}{}_b{}^\mu
\bar\theta^{-1}{}_c{}^\nu \bar\theta^{-1}{}_d{}^\rho \right)
\theta^a{}_\mu \theta^b{}_\nu \theta^c{}_\rho \theta^d{}_\sigma \cr
&+{1\over 2} \ell^{-2}\mu^{\prime2} \bkappa^{ac} \bkappa^{bd} 
\kappa_{ab} \kappa_{cd}
+ {1\over 4}\ell^{-4}\left(
 \lambda^\prime_1 \bkappa^{ac}\bkappa^{bd}\bkappa^{eg}\bkappa^{fh} 
+\lambda^\prime_2 \bkappa^{ah}\bkappa^{bc}\bkappa^{de}\bkappa^{fg}\right) 
                         \kappa_{ab} \kappa_{cd} \kappa_{ef} \kappa_{gh} 
\Bigr] \ . \cr} \eqno(22)$$
One could consider also other potential terms, for example the ones
given in equation (31), but these would not lead to any qualitatively new
feature.
Powers of $\ell$ have been inserted to compensate for the dimension of the
background fields (see the Appendix; these factors disappear when
we insert the explicit form of the background fields given below).
In this way $\mu$ has dimensions of mass and the coupling constants 
$\lambda$ are dimensionless.
For simplicity we are going to assume $\mu^\prime=\mu$, $\lambda^\prime_1=
\lambda_1$ and $\lambda^\prime_2=\lambda_2$.
In order to have nontrivial minima we are going to assume as usual that
$\mu^2<0$. 

As long as we allow $GL(4)$ and coordinate transformations
to act also on the background fields as in (5), 
the action (21) is invariant under the full gauge group discussed in section 2. 
Note that it is polynomial in the fields, containing 
at most quartic interactions. Thus, it is power-counting renormalizable.

In the Higgs model the ground state has the properties that $F_{\mu\nu}=0$
and $\nabla_\mu\Phi=0$. One can then choose the gauge so that $A_\mu=0$, and 
this implies that $\Phi$ is constant. 
The vacuum expectation value of $\Phi$ is obtained, up to a global
$G$ transformation, by minimizing the Higgs potential.
In the gravitational Higgs model the discussion is slightly modified
because the choice of the background fields breaks the gauge invariance
{\it ab initio}.

We choose $\btheta^a{}_\mu=\ell^{-1}\delta^a_\mu$ and
$\bkappa_{ab}=\ell^{-1}\eta_{ab}$, implying also 
$\bar g_{\mu\nu}=\eta_{\mu\nu}$.
These choices completely break the gauge invariance.
We assume that the energy functional arising from the terms 
$\bar S_2+\bar S_{\rm m}$ is minimized by the conditions
$$F_{\mu\nu}{}^a{}_b=\ 0\qquad ,\qquad
\itheta_\mu{}^a{}_\nu=\ 0\qquad ,\qquad
\idelta_{\lambda ab}=\ 0\ .\eqno(23)$$
Among all solutions of these equations we will consider only
$A=0$, $\kappa={\rm const}$ and $\theta={\rm const}$.
The vacuum expectation values $\langle\theta\rangle$ and 
$\langle\kappa\rangle$ are determined by minimizing the potential $U$. 
With the given background fields, the potential has absolute minima for
$$\langle\theta^a{}_\mu\rangle=
\sqrt{-{\mu^2\over4\lambda_1+\lambda_2}}\ \delta^a_\mu
\quad,\quad
\langle\kappa_{ab}\rangle=
\sqrt{-{\mu^2\over4\lambda_1+\lambda_2}}\ \eta_{ab}\ .\eqno(24)$$
Selfconsistency requires that the
vacuum expectation values of the fields coincide with the background fields.
This will be the case provided we identify
$$\ell=\sqrt{-{4\lambda_1+\lambda_2\over\mu^2}}\ .\eqno(25)$$
Since, as observed in section 3, the length $\ell$ is of the order of
Planck's length, the mass $\mu$ has to be of the order of Planck's mass.

Modulo the unproven assumption about the positivity of the energy,
we have thus found that when $\mu^2<0$ the ground state of the theory 
is Minkowski space. In the chosen gauge, it is given again by equation (17).
However, this time the nondegeneracy of the metric has 
been obtained as a result of the dynamics.
As in the previous section, expanding to second order around this ground state
will reveal the presence of a mass term for the gauge fields of the form (18).
Since now the gauge field propagates, a genuine Higgs phenomenon is
at work. Unfortunately, due to the presence of the potential,
also the fluctuations $\tilde\theta=\theta-\langle\theta\rangle$ and 
$\tilde\kappa=\kappa-\langle\kappa\rangle$ will have mass terms and
therefore, barring miracles, there will be no massless gravitons in 
this theory. It would seem that the price we pay for explaining the
nondegeneracy of $\theta$ and $\kappa$ is that the theory does not
describe gravity anymore.

I believe this model has to be regarded as a ``pregeometric''  model.
It describes the propagation of ``matter fields'' $\theta$,
$\kappa$ and $A$ in the background geometry provided by $\btheta$
and $\bkappa$. The coupling is $GL(4)$-- and generally covariant.
Then, quantum fluctuations of $\theta$, $\kappa$ and $A$ will generate
a $GL(4)$-- and generally covariant effective action for $\btheta$
and $\bkappa$. 
This action can be expanded in powers of derivatives of $\bar\theta$
and $\bar\kappa$ and at long wavelengths the dominant term will be 
the Hilbert action $S_H(\bar\theta,\bar\kappa)$.
All this is in complete analogy to usual ``induced gravity'' schemes [15]. 
Unlike most models of this type, however, in the present approach the
``matter fields'' are the unconstrained degrees of freedom of the 
gravitational field itself. In this sense, even though it seems to 
describe only flat space, the model presented here deserves to be
regarded as a model for gravity.

\beginsection 5. Unification

While the consequences of the Higgs point of view for quantum gravity are
still quite speculative, there are immediate dividends to be obtained 
in the field of unification [1]. In fact, a simple generalization of the model
presented in the previous section can be used to describe the unification 
of gravity with any Yang-Mills gauge theory, in the strict technical sense
in which the word is used in particle physics.

This generalization consists in replacing the 
vectorbundle $\xi_{\bR^4}$ by a larger vectorbundle $\xi_{\bR^N}$.
The dynamical variables are still $\theta^a{}_\mu$, $\kappa_{ab}$ and
$A_\lambda{}^a{}_b$, but now all latin indices are allowed to
run from $1$ to $N$, while spacetime 
indices $\mu,\nu,\ldots$ still run from $1$ to $4$. 
The discussion of section 2 can be repeated with some modifications.
The geometrical significance of the fields is unchanged, except that
now $\theta$ cannot be an isomorphism anymore: it is only a 
homomorphism from $TM$ to $\xi_{\bR^N}$. This is also reflected in the fact
that (1) cannot hold anymore; its generalization is the condition
$${\rm rank}\ \theta^a{}_\mu(x)=4 \quad\quad \forall x\ .\eqno(1^\prime)$$
Similarly, instead of (2) we now have 
$$\lambda_1(x)<0\ ,\ \lambda_2(x)>0\ , \ldots 
,\ \lambda_N(x)>0 \qquad\qquad\forall x\ .\eqno(2^\prime)$$
Given $\theta$, $\kappa$ and $A$ one can still form the induced metric and 
connection on spacetime. The induced metric is still defined by 
equation (3). However, equation (4) has to be modified since now 
$\theta^{-1}$ does not exist. The induced connection is now defined by
$$A_\lambda{}^\mu{}_\nu=
\ell^3\bigl(\theta_a{}^\mu A_\lambda{}^a{}_b \theta^b{}_\nu
+\theta_a{}^\mu \partial_\lambda \theta^a{}_\nu\bigr) ,\eqno(4^\prime)$$
where $\theta_a{}^\mu=g^{\mu\nu}\kappa_{ab}\theta^b{}_\nu$.
Note that $\theta^a{}_\mu \theta_a{}^\nu=\ell^{-3}\delta^\nu_\mu$ and
$\theta^a{}_\mu \theta_b{}^\mu=\ell^{-3}P^a{}_b$, where $P^a{}_b$ denotes a
projector onto the image of $\theta$ in $\xi_{\bR^N}$.

Assuming that ($1^\prime$) holds, the analog of the metric gauge
is now a gauge in which $\theta^a{}_\mu=\ell^{-1}\delta^a_\mu$ 
(with $\delta^a_\mu=0$ for $a>4$). In this gauge $\kappa_{\mu\nu}=
\ell^{-1}g_{\mu\nu}$ for $\mu,\nu=1,2,3,4$.
The residual gauge group consists
of coordinate transformations and local $G_{(N,4)}$ transformations,
where $G_{(N,4)}$ denotes the group of matrices of the form
$$\Lambda=\left[\matrix{{\bf 1}_4&a\cr0&b\cr}\right] \eqno(26)$$
with $b\in GL(N-4)$. 

If ($2^\prime$) holds, the $N$-bein gauge is 
defined by the condition $\kappa_{ab}=\ell^{-1}\eta_{ab}$; 
in this gauge local $GL(N)$ invariance is broken to
local $O(1,N-1)$ invariance. 

It is still true that the two gauge conditions cannot be imposed simultaneously.
However, using the residual $G_{(N,4)}$ gauge freedom of the metric gauge,
one can further impose that $\kappa_{ab}=\ell^{-1}\delta_{ab}$ 
for $a,b=5,\ldots,N$
and $\kappa_{ab}=0$ for $a=1,2,3,4$ and $b=5,\ldots,N$ or vice-versa.
So altogether 
$$\theta=\ell^{-1}\left[\matrix{{\bf 1}_4\cr 0\cr}\right]\qquad,\qquad
\kappa=\ell^{-1}\left[\matrix{g & 0 \cr
0 & {\bf 1}_{N-4} \cr}\right] \eqno(27)$$
We will call this the extended metric gauge. In this gauge local 
$GL(N)$ invariance is broken down to local $O(N-4)$ invariance.

If $\theta$ and $\kappa$ have maximal ranks, the internal space 
at each point $x$ can be split into $\xi_x={\rm im}\,\theta(x)\oplus\zeta_x$,
where ${\rm im}\,\theta(x)$ denotes the image of $\theta$ and $\zeta$
is the orthogonal complement. Accordingly, the gauge field 
$A_\lambda{}^a{}_b$ can be decomposed naturally in four parts:
$$A_\lambda=\left[\matrix{A_\lambda^{(4)} & H_\lambda \cr
K_\lambda & A_\lambda^{(N-4)} \cr}\right]\ .\eqno(28)$$
In the extended metric gauge, the components $A_\lambda^{(4)}$ can be
identified with the components of the induced connection on spacetime
($4^\prime$), the components $A_\lambda^{(N-4)}$ represent a purely
internal $GL(N-4)$ Yang-Mills field and the off-diagonal components
$H_\lambda$ and $K_\lambda$ correspond to a mixing of internal and spacetime
variables. In this way the gravitational connection, carrying a representation
of $GL(4)$, and a Yang-Mills field, carrying a representation of
$GL(N-4)$ have been put together into a bigger gauge field, carrying
a representation of $GL(N)$. This act of putting together the representations
of two groups into a representation of a larger group is the essence
of unification, in the sense in which the word is used in particle
physics. 

This splitting of the $GL(N)$ connection is only possible if $\theta$
has maximal rank. At the other extreme, when $\theta$ is identically zero,
all directions in the internal spaces are equivalent.
We thus come to the important conclusion
that the soldering form is the order parameter whose vacuum
expectation value tells us whether the symmetry between the
gravitational and nongravitational interactions is broken.
(Since a nondegenerate $g_{\mu\nu}$ implies equation ($1^\prime$),
one can also say that the order parameter is the spacetime metric.
Purists may prefer this since $g_{\mu\nu}$ is invariant under local $GL(N)$
transformations).

There is another aspect of the discussion in section 2 which needs modification
in the case $N>4$. The definitions (6) and (7) of torsion and nonmetricity
remain unchanged, but now the equations $\itheta=0$ and $\idelta=0$
do not have a unique solution. Assuming that ($1^\prime$) and
($2^\prime$) are fulfilled, the solutions
are those gauge fields such that, in the extended metric gauge, 
$A^{(4)}{}_\lambda{}^\mu{}_\nu=
\igamma_\lambda{}^\mu{}_\nu$ (the Christoffel symbols of $g$),
$A^{(N-4)}_{\lambda ab}=-A^{(N-4)}_{\lambda ba}$,
$H_{\lambda ab}=-K_{\lambda ba}$ and
$K_\lambda{}^a{}_b\theta^b{}_\tau=K_\tau{}^a{}_b\theta^b{}_\lambda$.
Thus when $\theta$ and $\kappa$ have maximal ranks, the solutions
of the equations $\itheta=0$ and $\idelta=0$ depend on $2N(N-4)$
arbitrary functions [16].

A consequence of this is that if we were to repeat the analysis
of section 3 in the case $N>4$, the equations of motion would not determine
$A$ completely. For this and other reasons which will become apparent 
later, it is preferable to replace the action $S_{\rm m}$ given in (10)
by another action containing instead of $\itheta_\mu{}^a{}_\nu$ the
covariant derivative
$$\nabla_\mu\theta^a{}_\nu=\partial_\mu\theta^a{}_\nu
+A_\mu{}^a{}_b\theta^b{}_\nu
-\igamma_\mu{}^\lambda{}_\nu\theta^a{}_\lambda\ .\eqno(29)$$
Note that if $\nabla_\mu\theta^a{}_\nu=0$, the induced
connection ($4^\prime$) is given by the Christoffel symbols of $g$.
Note also that 
$\itheta_\mu{}^a{}_\nu=\nabla_\mu\theta^a{}_\nu-\nabla_\nu\theta^a{}_\mu$.
Since the covariant derivative has no symmetry properties in the indices,
there are many possible terms quadratic in $\idelta$ and $\nabla\theta$.
We will not write here the most general expression. 

At this point it should be clear how the gravitational Higgs model of the 
previous section has to be generalized in the case $N>4$.
We will use again a selfconsistent method with an action of the form (21),
except for the following modifications. 
In $S_{\rm m}$ we will replace the term quadratic in $\itheta$ 
by terms quadratic in $\nabla\theta$. 
Actually, since the Christoffel symbols $\igamma_\lambda{}^\mu{}_\nu$ 
appearing in (29) are nonpolynomial in $\kappa$ and $\theta$, 
in the selfconsistent scheme they have to be replaced by the Christoffel 
symbols $\bar\igamma_\lambda{}^\mu{}_\nu$
of the background metric $\bar g$. When $\igamma$ is replaced by
$\bar\igamma$ in (29), the corresponding covariant derivative will be
denoted $\bar\nabla\theta$. As observed before, there are many
possible ways of contracting two covariant derivatives of $\theta$
but for the purpose of illustration it will be sufficient to consider
the following terms:
$$\eqalign{\bar S_{\rm m}(\theta,\kappa,A;\btheta,\bkappa)=
{1\over2}\int d^4x \sqrt{|\det \bar g|}\,&
\bigl[A_1\ell\,\bar g^{\mu\rho}\bar g^{\nu\sigma}\bar\kappa_{ab}
\bar\nabla_\mu\theta^a{}_\nu \bar\nabla_\rho\theta^b{}_\sigma \cr
&+B_1\ell^{-2}\bar g^{\mu\nu}\bar\kappa^{ac}\bar\kappa^{bd}
\nabla_\mu\kappa_{ab}\nabla_\nu\kappa_{cd}
\bigr]\ .} \eqno(30)$$
In the potential $U$ given in equation (22), 
$\ell^{-1}\bar\theta^{-1}{}_a{}^\mu$
has to be replaced by $\ell^2\bar\theta_a{}^\mu$. 
We then observe that with the background fields given below,
this potential does not depend on the components $\theta^a{}_\mu$
with $a>4$. Since we want the potential to determine the fields 
as much as possible, we add to $U$ further terms
$$\int d^4x \sqrt{|\det \bar g|}\Bigl[
{1\over 2} \ell\mu^2\bar g^{\mu\nu}\bkappa_{ab}\theta^a{}_\mu \theta^b{}_\nu
+{1\over 4}\bar g^{\mu\nu} \bar g^{\rho\sigma} \ell^2
\left(\lambda_1 \bkappa_{ab} \bkappa_{cd}+\lambda_2 \bkappa_{ad} \bkappa_{bc} 
\right)\theta^a{}_\mu \theta^b{}_\nu \theta^c{}_\rho\theta^d{}_\sigma \Bigr]\ .
\eqno(31)$$
We now choose the background fields $\btheta^a{}_\mu=\ell^{-1}\delta^a_\mu$, 
$\bkappa_{ab}=\ell^{-1}\eta_{ab}$, $\bar g_{\mu\nu}=\eta_{\mu\nu}$.
This choice breaks local $GL(N)$ invariance down to local $O(N-4)$
invariance. Having fixed the background fields, the potential for $\kappa$
given in the last line of (22) and the potential for $\theta$ given in (31)
remain invariant under local $O(1,N-1)$ transformations, while the potential
for $\theta$ given in (22) remains invariant under local $GL(N-4)$ 
transformations.
Altogether the potential, and hence the whole action $S(\theta,\kappa,A;
\bar\theta,\bar\kappa)$, remains invariant under local 
$O(N-4)$ transformations.

The analysis of the previous section can be repeated and gives for the
ground state $\langle A_\mu{}^a{}_b\rangle=0$, 
$\langle\theta^a{}_\mu\rangle=
\sqrt{-{\mu^2\over4\lambda_1+\lambda_2}}\, \delta^a_\mu$ 
and 
$\langle\kappa_{ab}\rangle=\sqrt{-{\mu^{\prime^2}
\over N\lambda_1^\prime+\lambda_2^\prime}}\ \eta_{ab}$. 
This corresponds again to flat space and selfconsistency is achieved again 
by choosing $\ell$ as in (25), provided also 
${N\lambda^\prime_1+\lambda^\prime_2\over \mu^{\prime2}}=
{4\lambda_1+\lambda_2\over \mu^{2}}$.

When the action is expanded to second order around this ground state,
$\bar S_{\rm m}$ gives rise to a term of the type
$$\eqalign{{1\over2}\ell^{-2}\!\int d^4x\, 
\bigl[A_1\eta^{\mu\nu}A_\mu{}^a{}_c A_\nu{}^b{}_d \, 
\eta_{ab}P^c{}_fP^d{}_g\eta^{fg}
+B_1\eta^{\mu\nu}\eta^{ac}\eta^{bd}
(A_{\mu ab}+A_{\mu ba})(A_{\nu cd}+A_{\nu dc})\bigr] \ .\cr}\eqno(32)$$
In order to diagonalize the mass matrix we define the combinations
$$\eqalignno{
A^{(4,\pm)}_{\lambda ab}=&\ 
{1\over2}(A^{(4)}_{\lambda ab}\pm A^{(4)}_{\lambda ba})\ ,&(33a)\cr
A^{(N-4,\pm)}_{\lambda ab}=&\ 
{1\over2}(A^{(N-4)}_{\lambda ab}\pm A^{(N-4)}_{\lambda ba})
\ ,&(33b)\cr
H^{(\pm)}_{\lambda ab}=&
{1\over\sqrt2}
\left(\sqrt{1\mp{A_1\over\sqrt{A_1^2+4B_1^2}}}\ H_{\lambda ab}
\pm
\sqrt{1\pm{A_1\over\sqrt{A_1^2+4B_1^2}}}K_{\lambda ba}\ \right)
\ ,&(33c)\cr}$$
where the internal indices are restricted to the appropriate ranges.
Then, (32) can be rewritten
$$\eqalign{&{1\over2}\ell^{-2}\!\int\! d^4x \,\eta^{\mu\nu} 
\Bigl[(A_1\!+\!4B_1) A^{(4,+)}_{\mu ab} A^{(4,+)}_\nu{}^{ab}
\!+\!A_1 A^{(4,-)}_{\mu ab} A^{(4,-)}_\nu{}^{ab}
\!+\!4B_1 A^{(N-4,+)}_{\mu ab} A^{(N-4,+)}_\nu{}^{ab} \cr
&+\!{1\over2}\!\left(A_1\!+\!2B_1\!+\!\sqrt{A_1^2\!+\!4B_1^2} \right)\!
H^{(+)}_{\mu ab} H^{(+)}_{\nu}{}^{ab}\!
+\!{1\over2}\!\left(A_1\!+\!2B_1\!-\!\sqrt{A_1^2\!+\!4B_1^2}\right)\!
H^{(-)}_{\mu ab} H^{(-)}_{\nu}{}^{ab}\Bigr] .\cr}\eqno(34)$$
We see that the only massless part of the connection is 
$A^{(N-4,-)}$, describing a ``purely internal'' $O(N-4)$ Yang-Mills field.
All other fields have acquired a mass of the order of Planck's mass.
The fact that the components $A^{(4)}$ of the gauge field acquire
a large mass corresponds to the observed fact that gravity is not mediated
by massless spin one fields. The mixed components $H^{(\pm)}$
which are needed to complete the $GL(N)$ multiplet also naturally
become massive. The mass of $A^{(N-4,+)}$ breaks $GL(N-4)$ to
$O(N-4)$.  

It is interesting to consider the limit $B_1\rightarrow\infty$.
In this case $H^{(+)}$ becomes the symmetric combination of $H$
and $K$ and its mass diverges, whereas $H^{(-)}$ becomes the antisymmetric
combination of $H$ and $K$ and its mass becomes $\sqrt{A_1/2}$.
This limit corresponds to imposing $\idelta=0$ as a constraint.

Finally, we note that for $N=14$ the unbroken group $O(10)$ can be
immediately identified with the grand unification group.

\beginsection 6. Conclusions

We have seen that a Higgs mechanism is operative in theories of gravity in
which a connection is a dynamical variable. In some sense this is true
already in GR, and much more so in theories with Lagrangians quadratic
in curvature. However, as long as the constraints (1) and (2) are
imposed by hand, these theories are really closer to gauged nonlinear 
sigma models than gauged Higgs models. I have proposed a true Higgs-like 
model for gravity, in which the conditions (1) and (2) arise as properties of
the ground state rather than {\it a priori} constraints. The problem of 
writing an action for a theory with possibly degenerate metric was circumvented
by postulating a selfconsistent dynamics, in which the dynamical fields
appear in the action at most quartically and are coupled to their own
mean values (assumed nondegenerate) which appear nonpolynomially.

There has been much speculation recently about the possible existence of a
``topological'', metric-less phase of gravity [12,17]. However, none of these
approaches has provided a dynamical explanation for the nondegeneracy
of the (vacuum expectation value of the) metric. The Higgs-like model
proposed here is a first attempt in this direction.

The other purpose of the model is to provide a framework for the unification 
of gravity with the other interactions. Superficially, it may seem to
bear some resemblance to Kaluza-Klein theories. However, the two
mechanisms are profoundly different. In Kaluza-Klein theories, spacetime
has more than four dimensions and some components of the metric in the 
extra dimensions are reinterpreted as Yang-Mills fields.
Here, spacetime remains four dimensional. Instead, the internal spaces are
enlarged and some of the components of the internal metric and connection
are reinterpreted as spacetime metric and connection. Thus in
Kaluza-Klein theories ``spacetime structures are converted to internal
structures'' while here ``internal structures are converted to
spacetime structures''. This approach to unification is also not new:
it had been suggested by Einstein and Meyer in 1931 as a way of unifying
gravity and electromagnetism [18]. What is new here is the recognition of the 
soldering form as an order parameter, and the dynamical mechanism for
symmetry breaking.  I believe that in this updated version this is the most 
natural approach to unification from the point of view of a particle physicist.

A rather unfamiliar feature of this approach to unification is the
vectorial nature of the order parameter. Usually, in the Higgs phenomenon,
the order parameter is a scalar field. Let us note in this connection that a
nonzero vacuum expectation value of the soldering form does not break
Lorentz invariance. In fact, while the ``internal'' and ``spacetime''
Lorentz groups, acting on $\theta$ as $\theta\rightarrow \Lambda^{-1}\theta$
and $\theta\rightarrow \theta L$ are indeed broken by the choice
$\theta^a{}_\mu=\delta^a{}_\mu$, the ``diagonal'' subgroup
$\theta\rightarrow L^{-1}\theta L$ remains unbroken.

Throughout this paper I have strived to treat $\theta$ and $\kappa$ 
on an equal footing. However, it appears from the discussion in section 6
that the crucial role in this unification scheme is played by $\theta$. 
There is also another difference which is worth mentioning: 
while the forbidden region defined by
the constraint (1) (or ($1^\prime$)) is nowhere dense in the space of the 
fields, 
the one defined by (2) (or ($2^\prime$)) is open. 
Therefore, the constraint (2) (or ($2^\prime$)) has better chances
of being implementable at the quantum level.  
It appears from these remarks that the introduction of the fiber metric
$\kappa$ as a separate variable, while useful at this stage,
may not be necessary in the construction of a realistic model.
As a first try, it is natural to consider a model
in which $\kappa_{ab}=\ell^{-1}\eta_{ab}$ from the outset [1,19].

The construction of a realistic grand unified theory will require also the
introduction of spinor fields. Here there are various possibilities.
If the theory can be consistently quantized with the constraint ($2^\prime$)
as suggested above, then one could start with $O(N)$ (iso)spinors,
and when soldering occurs ({\it i.e.} when $\theta$ acquires a nonzero
expectation value) these become world spinors and $O(N-4)$ isospinors.
This is precisely the kind of matter fields that appear in $O(10)$ grand
unified theories. If on the other hand also the fiber metric $\kappa$ is 
allowed to become degenerate, then ordinary spinors cannot be fundamental
fields. To my knowledge, there are two possible solutions.
The first is to use infinite dimensional spinorial representations
of $GL(N)$, as suggested in [20]. The other is to use so-called 
``K\"ahler fermions'' [21]. These are fields with values in the Grassmann 
algebra of differential forms. In the broken phase, when the metric 
becomes nondegenerate, the Grassmann algebra can be endowed with a Clifford
structure and the fields then obey Dirac's equation.

What will be the consequences of our point of view on quantum gravity?
As stated in section 5, the overall picture should be very similar
to the one occurring in so-called ``induced gravity'' schemes,
but with some additional peculiarities.

Since the quantization of fields in flat space can be done preserving
the Lorentz symmetry, when the models of sections 4 and 5 will be quantized, 
the ground state will still correspond to Minkowski space, 
except perhaps for an overall scaling of the metric [19].
In fact, since the metric $g$ is an order parameter, its 
vacuum expectation value could depend upon the scale at which the system is
observed. This is highly reminiscent of Dirac's idea that the metric governing
atomic phenomena may be different from the macroscopic metric [22].
It is conceivable that the expectation value of the metric will
decrease towards shorter distances and vanish at Planck's length. 
One is then tempted to speculate that particles with higher
momentum will see a smaller metric, the net effect being a cutoff 
on momentum integrations at Planck's mass. In this way the ``Higgs picture''
may provide an implementation of the old idea that gravity acts
as a universal regulator [23]. 

Critics say that the Higgs mechanism is only an {\it ad hoc} construction
and should not be taken as a description of nature at the fundamental
level. Instead, the explanation of the origin of
masses should be looked for in the domain of nonperturbative phenomena.
Whether fundamental or not, the Higgs mechanism has been undeniably
very successful in categorizing the raw data of particle physics.
I hope the ideas presented here can be equally useful in bridging the gap
between the world of particle physics and that of gravitational phenomena.

\bigskip
\centerline{\bf Acknowledgements}
\noindent I wish to thank R. Floreanini and E. Spallucci for many 
useful discussions.

\vfil\eject

\beginsection Appendix. Dimensional assignments

In theories of gravity there are conflicting requirements on the dimensions 
to be assigned to the fields, coming from their dual roles as geometric
standards and dynamical variables. This forces the introduction of a 
fundamental dimensionful constant $\ell$. 

The dimension of a quantity $Q$, denoted $d(Q)$, is defined as follows:
when units of length (abbreviated $L$) scale by a factor $\alpha$,
$Q$ scales by a factor $\alpha^{d(Q)}$. We will also say that $Q$
has dimension $L^{d(Q)}$.

We begin by considering the line element $ds^2=g_{\mu\nu}dx^\mu dx^\nu$.
The geometrical interpretation requires that its dimension be 2.
One can then choose arbitrarily the dimension of the coordinates, $d(x)$, 
and the dimension of the metric will be $d(g_{\mu\nu})=2(1-d(x))$. 
We will consider here only the choices $d(x)=1$, $d(g_{\mu\nu})=0$ 
and $d(x)=0$, $d(g_{\mu\nu})=2$.
The second choice is the most natural geometrically, since the coordinates
are then mere numbers labelling spacetime points. It is also more natural
in a theory in which the metric tensor is allowed to become zero.
However, the first choice is more familiar and also has the advantage that
all fields have the same canonical dimensions.
Therefore, in the text we have assumed that the coordinates have dimension 1.
 
The canonical dimension of a bosonic field 
$T_{\mu_1\ldots\mu_r}{}^{\nu_1\ldots\nu_s}$
in $n$ spacetime dimensions can be computed from the kinetic term
$$-{1\over2}\int d^nx\ \sqrt{|\det g|}\
g^{\lambda\tau}g^{\mu_1\rho_1}\ldots g^{\mu_r\rho_r}
g_{\nu_1\sigma_1}\ldots g_{\nu_s\sigma_s}
\partial_\lambda T_{\mu_1\ldots\mu_r}{}^{\nu_1\ldots\nu_s}
\partial_\tau T_{\rho_1\ldots\rho_r}{}^{\sigma_1\ldots\sigma_s} \eqno(A.1)$$
We see that if $d(x)=1$, $d(T)={2-n\over2}$, whereas if $d(x)=0$,
$d(T)={2-n\over2}+r-s$. 
In order to discuss quantities with internal indices, one has to decide
first what dimensions to assign to the fiber metric $\kappa$.
As noted in the text, $\kappa$ is a nonlinear sigma model with values in
$GL(4)/O(1,3)$; therefore, its action will be nonpolynomial.
The typical action for such a field $\tilde\kappa$ is
$$S=-{B\over2}\int d^nx\ \sqrt{|\det g|}\ g^{\mu\nu}
\partial_\mu\tilde\kappa_{ab}\,\partial_\nu\tilde\kappa_{cd}\,
h^{abcd}(\tilde\kappa)\ ,\eqno(A.2)$$
where for example $h^{abcd}(\tilde\kappa)=\tilde\kappa^{ac}\tilde\kappa^{bd}$
and $\tilde\kappa^{ac}\tilde\kappa_{cb}=\delta^a_b$.
This form of the action says nothing on the dimension of the scalar fields.
Here $\tilde\kappa$ has been taken dimensionless and $d(B)=2-n$. 
The canonical dimension is determined by the perturbative kinetic term
which is obtained by defining $\kappa_{ab}=\sqrt{B}\, \tilde\kappa_{ab}$
and replacing $\kappa$ by its background value $\bar\kappa$ in
$h^{abcd}$:
$$S=-{1\over2}\int d^nx\ \sqrt{|\det g|}\ g^{\mu\nu}
\partial_\mu\kappa_{ab}\,\partial_\nu\kappa_{cd}\,
h^{abcd}\left({\bar\kappa\over\sqrt{B}}\right)
+{\rm interaction\ terms}\ .\eqno(A.3)$$
In this way we find $d(\kappa)={2-n\over2}$, as for scalars without internal
indices. In the same way, if the tensor $T$ carries internal indices,
in the perturbative kinetic term these have to be contracted with a
dimensionless background fiber metric, {\it e.g.} ${\bar\kappa\over\sqrt{B}}$.
Therefore, the canonical dimensions of the fields are independent of the
number of internal indices.

From now on we consider only the case $n=4$ and $d(x)=1$.
In this case the fields $\theta^a{}_\mu$, $\kappa_{ab}$ and $A_\mu{}^a{}_b$
have canonical dimension $-1$.

One would try to define $g_{\mu\nu}=\theta^a{}_\mu\theta^b{}_\nu\kappa_{ab}$,
but this object does not have the right dimensions for a metric.
So one is forced to introduce a fundamental length $\ell$ and define
$g_{\mu\nu}$ as in (3) (note that actually the power of $\ell$ appearing in (3)
is independent of the choice of $d(x)$).
The constant $\ell$ is to be regarded as a fundamental constant of Nature,
like the velocity of light and Planck's constant. Once introduced, it
can be used freely in the construction of the Lagrangian.
For example, the Palatini action (9) has been written in the familiar form 
in which $G$ can be identified with Newton's constant, 
but if we had tried to write the action using only the fundamental fields
and the constant $\ell$ we would have arrived at 
$$S_P=\ell\int d^4x\ \sqrt{|\det \kappa|}|\det\theta|\
\theta^{-1}{}_a{}^\mu\theta^{-1}{}_c{}^\nu\kappa^{bc} 
F_{\mu\nu}{}^a{}_b\ .\eqno(A.4)$$
If we use the definition (3) in (9) and compare with (A.4), 
we see that $\ell=\sqrt{16\pi G}$; thus, not unexpectedly, 
the fundamental length must be of the order of Planck's length.

The raising and lowering of spacetime indices is performed by means
of the dimensionless spacetime metric. It is convenient to perform also
other tensorial operations by means of dimensionless geometric objects.
For instance, spacetime indices can be transformed to internal indices using
the dimensionless soldering form $\ell\theta^a{}_\mu$ and
internal indices can be raised and lowered by means of the 
dimensionless fiber metric $\ell\kappa_{ab}$. In this way the dimension
of a tensor is independent of the nature and position of its indices.
This is the convention that I have adopted in writing equations
(10,11,15,16,19,20,22). In this way the coefficients $A_i$, $B_i$, $C_i$,
$G_i$ of the kinetic terms are all dimensionless, the coefficients
$\ell^{-2}q_i$ and $\mu^2$ of mass terms have dimensions $-2$ 
and the coefficients
$\lambda$ of the quartic interaction terms are dimensionless.

Finally we record here the relation between the coefficients 
$A_i$, $B_i$, $C_i$ appearing in (11) and the coefficients $q_i$ 
appearing in (16):
$$\eqalign{&q_1=2A_1+2B_1+2C_1 \cr 
&q_3=-2A_1+B_2-2C_1 \cr
&q_5=-2A_2+2B_2-2C_1-{\ell^2\over8\pi G} \cr
&q_7=B_4 \cr
&q_9=2B_5-2C_5 \phantom{1\over2}\cr
&q_{11}=-2A_3+2B_5+4C_2-2C_3 \cr}\qquad
\eqalign{&q_2=A_2+2B_1+2C_1 \cr
&q_4=A_2+B_2 \cr
&q_6=A_3+4B_3-4C_2 \phantom{1\over2} \cr 
&q_8=A_3+B_4+2C_3 \cr
&q_{10}=2B_4+2C_3+{\ell^2\over8\pi G} \cr
&\phantom{q_{11}=-2A_3+2B_5+4C_2-2C_3} \cr}\eqno(A.5)$$

\vfil\eject

\centerline{\bf References}
\bigskip
\noindent
\item{1.}R. Percacci; Phys. Lett. {\bf 144 B}, 37 (1984).
\item{2.}R. Percacci; in ``XIII Conference on Differential
Geometric Methods in Theoretical Physics'', World Scientific (1984).
\item{3.}R. Percacci; ``Geometry of Nonlinear Field Theories'',
World Scientific (1986).
\item{4.}A. Komar; Phys. Rev. {\bf D 30} 305, (1984); 
J. Math. Phys. {\bf 26}, 831 (1985);\hfil\break
Y. Ne'eman and Dj. Sijacki; Phys. Lett {\bf B 200},
489 (1988).
\item{5.}R. Floreanini and R. Percacci; Class. and Quantum Grav. {\bf 7}, 
975 (1990). 
\item{6.}R. Floreanini and R. Percacci; Class. and Quantum Grav. {\bf 7}, 
1805 (1990).
\item{7.}R. Floreanini and R. Percacci; ``Topological $GL(3)$-invariant
gravity'', preprint SISSA 97/90/EP, to appear in Class and Quantum 
Grav.
\item{8.} C. Isham, A. Salam and J. Strathdee; Ann. of Physics {\bf 62}, 98
(1971);\hfil\break
K. Cahill; Phys. Rev. {\bf D 18}, 2930 (1978);\hfil\break
D. Popovi\'c; Phys. Rev. {\bf D 34},1764 (1986);\hfil\break
J. Dell, J.L. deLyra and L. Smolin; Phys. Rev. {\bf D 34}, 3012 (1986).
\item{9.}E. Sezgin and P. van Nieuwenhuizen; Phys. Rev. {\bf D 21}, 3269 (1980);
\hfil\break
E. Sezgin; Phys. Rev. {\bf D 24}, 1677 (1981);\hfil\break
K. Hayashi and T. Shirafuji; Prog. Theor. Phys. {\bf 64}, 866 (1980);
{\it ibid.} 883 (1980); {\it ibid.} 1435 (1980); {\it ibid.} 2222 (1980);
\hfil\break
R. Kuhfuss and J. Nitsch; Gen. Rel. and Grav. {\bf 18}, 1207 (1986).
\item{10.}L. Smolin; Nucl. Phys. {\bf B 247}, 511 (1984).
\item{11.}A. Ashtekar; ``New perspectives in canonical gravity'',
Bibliopolis, (1988).
\item{12.}R. Capovilla, T. Jacobson and J. Dell;
Phys. Rev. Lett. {\bf 63}, 2325 (1990).
\item{13.}I. Bengtsson; Classical and Quantum Grav. {\bf 7}, 27 (1990).
\item{14.}C. Isham and A.C. Kakas; Classical and Quantum Grav. {\bf 1},
621 (1984); {\it ibidem} 633.
\item{15.}S. Adler; Rev. Mod. Phys. {\bf 54}, 729 (1982);\hfil\break
D. Amati and G. Veneziano; Nucl. Phys. {\bf B 204}, 451 (1982).
\item{16.}R. Percacci, in ``Fibre bundles: their use in physics'',
World Scientific, Singapore (1988).
\item{17.}E. Witten; Comm. Math. Phys. {\bf 117}, 353 (1988);
Phys. Lett. {\bf 206 B}, 601 (1988).
\item{18.}A. Einstein and M. Meyer; Sitzungsber. der Preuss. Akad.
der Wiss. (1931) p. 541; {\it ibid} (1932) p. 130.
\item{19.}R. Floreanini, R. Percacci and E. Spallucci;
``Why is the metric nondegenerate?'', preprint SISSA 132/90/EP.
\item{20.}Y. Ne'eman and Dj. Sijacki; Phys. Lett. {\bf 157 B}, 267, 
275 (1985).
\item{21.}I.M. Benn and R.W. Tucker; Comm. Math. Phys. {\bf 89}, 341 (1983).
\item{22.}P.A.M. Dirac; Proc. Roy. Soc. Lond {\bf A333}, 403 (1973).
\item{23.}S. Deser; Rev. Mod. Phys. {\bf 29}, 417 (1957).

\break
\centerline{\bf Postscript, december 2007}
\bigskip
Several developments have taken place in the last few years.
There is now a number of papers that discuss a form of
``Gravitational Higgs Phenomenon'' (GHP) which is different
from the one that was introduced here. Some representative references are
\smallskip
\item{24.} N. Arkani-Hamed, H. Georgi, M.D. Schwartz, Ann. Phys. {\bf 305} 96 (2003), 
\item{25.} Z. Berezhiani, D. Comelli , F. Nesti, L. Pilo, Phys.Rev.Lett.99:131101,2007.
\item{26.} G.'t Hooft, e-Print: arXiv:0708.3184 [hep-th]
\item{27.} Z. Kakushadze, e-Print: arXiv:0710.1061 [hep-th]
\item{28.} M. Maeno and I. Oda,  e-Print: arXiv:0801.0827 [hep-th]
\smallskip
\noindent
To clarify the distinction, let me call (for want of a better name) ``High Energy GHP''
the one discussed in this paper and ``Low Energy GHP'' the one
discussed in the papers mentioned above.
The following table highlights the differences between the two GHP:
\bigskip
{
\offinterlineskip
\tabskip=0pt
\halign{ 
\vrule height2.75ex depth1.25ex width 1pt 
#\tabskip=1em &\hfil#\hfil &\vrule width 0.6pt# & \hfil#\hfil 
&\vrule width 0.6pt# &
\hfil #\hfil &#\vrule width 0.6pt \tabskip=0pt &\hfil #\hfil\vrule width 0.6pt&\hfil #\hfil\vrule width 1pt\cr
\noalign{\hrule height 1pt}
& && scale && order parameter &&  breaks &\ \ gives mass to\ \ \cr
\noalign{\hrule}
& HEGHP && Planck && metric, soldering form && \ \ frame rotations\ \ & connection\cr
& LEGHP && Hubble && fluid/scalars &&\ \ diffeomorphisms\ \ & graviton\cr
\noalign{\hrule height 1pt}
}}
\bigskip
The HEGHP explains why gravity behaves at sub--Planckian energies
in a way that so closely resembles
chiral perturbation theory. It is a very compelling way of interpreting
gravity from the point of view of particle physics.
The LEGHP is for now an interesting speculation (it may have applications also
outside the gravitational context, e.g. in strong interaction physics).
It is worth noting that the interpretation adopted here of the vierbein/soldering as a map
between two spaces, rather than a frame in one space,
is very close in spirit to the two-site model discussed in [24].

Other recent references that address the HEGHP are
\smallskip
\item{29.} I. Kirsch, Phys. Rev. {\bf D 72} 024001 (2005), e-Print: hep-th/0503024
\item{30.} M. Leclerc, Annals Phys. {\bf 321} 708-743 (2006), e-Print: gr-qc/0502005
\smallskip
\noindent An interesting discussion of gravitational Nambu-Goldstone fields
in the context of Lorentz--violating theories, and the associated
Higgs phenomenon, can be found in
\smallskip
\item{31.} Robert Bluhm , V.Alan Kostelecky, Phys.Rev.D71:065008,2005.
\smallskip

The unification scheme based on the HEGHP 
has been recently reexamined in some more detail in 
\smallskip
\item{32.} F. Nesti and R. Percacci, e-Print: arXiv:0706.3307 [hep-th], to appear in J. Phys. A
\item{33.} S. Alexander, e-Print: arXiv:0706.4481 [hep-th]
\smallskip
\noindent The ``exceptionally simple'' $E_8$ unification of 
\smallskip
\item{34.} A.G. Lisi, e-Print: arXiv:0711.0770 [hep-th]
\smallskip
\noindent belongs to the same class of models, where
the gravitational and Yang--Mills connections are considered as
parts of the connection of a larger group. 
Lisi tries to go further by putting also all other fields in 
the same representation; on the other hand, he does not discuss
the $E_8$-breaking mechanism, not even at the kinematical level.
Motivated by [34],
\smallskip
\item{35.} L.Smolin, e-Print: arXiv:0712.0977 [hep-th]
\smallskip
\noindent has proposed a generalization of the Plebanski action 
that could be used in these unified theories. 
As an example he applies it to a model with gauge group $SO(8)$.
This action admits solutions that describe either the
symmetric or the broken phase of the theory.

\bye